# Revival of Superconductivity by $Y^{3+}/Ca^{2+}$ substitution in $YBa_2Cu_{2.7}Co_{0.3}O_7$ without reported phase transformation

M.A. Ansari, Rashmi Nigam, V.P.S. Awana*,$, Anurag Gupta, R.B. Saxena and H. Kishan

National Physical Laboratory K.S. Krishnan Marg, New Delhi 110012, India

N.P. Lalla, V. Ganesan and A.V. Narlikar

Inter-University Consortium for DAE Facilities, University Campus, Khandwa Road, Indore-452017, MP, India

C A. Cardoso$

Center for Superconducity Research, University of Maryland, College Park, MD 20742-4111, USA

Results of phase formation, resistivity ($\rho$), and thermo-electric power (S), are reported on $Y_{1-x}Ca_xBa_2Cu_{2.7}Co_{0.3}O_7$ compounds with x = 0.1 and 0.2. Pristine compound i.e. without Co or Ca substitution crystallizes in orthorhombic structure with space group P/mmm. Cu-site Co substituted compound i.e. $YBa_2Cu_{2.7}Co_{0.3}O_7$ is tetragonal. With simultaneous doping of Ca at Y site in Co substituted compound i.e. $Y_{1-x}Ca_xBa_2Cu_{2.7}Co_{0.3}O_7$ the tetragonal nature still remains. $\rho(T)$ measurements showed superconducting transition temperature ($T_c$) to decrease from 90K ($YBa_2Cu_3O_7$) to 33 K for $YBa_2Cu_{2.7}Co_{0.3}O_7$ which with further Ca substitution increases from 33K to 53K ($Y_{0.9}Ca_{0.1}Ba_2Cu_{2.7}Co_{0.3}O_7$) and 67 K for $Y_{0.8}Ca_{0.2}Ba_2Cu_{2.7}Co_{0.3}O_7$. $T_c$ decreases first with Cu-site Co substitution by hole-filling and later recovers by simultaneous hole creation by Y site Ca substitution. Room temperature thermoelectric power S(300 K), which is an indirect measure of mobile carriers shows the decrease of carriers with Co doping and creation by Ca substitution. Our results demonstrate the hole filling by Co substitution is compensated by simultaneous Ca substitution.

Key Words: $Y_{1-x}Ca_xBa_2Cu_{2.7}Co_{0.3}O_7$ compounds, Thermoelectric power, Hole filling and hole creation.

* Corresponding Authors: e-mail; awana@mail.nplindia.ernet.in

$ Presenting Author

## INTRODUCTION

Various on-site substitutional studies in high $T_c$ superconducting (HTSc) compounds had attracted a lot of attention, for example [1] and references there in. Basically all HTSc compounds in their ground state are antiferromagnetic insulators with Cu spins ordering above room temperature [1]. By doping of carriers through charge neutrality with various on-site alliovalent substitutions or oxygen content, one frustrates the Cu magnetic ordering and brings in the metallic behavior accompanied with superconductivity at low temperatures [1]. Phase diagrams are drawn in terms of doped carriers on the basis of mentioned substitutional studies for different HTSc families [1,2].

In a p-type (most HTSc compounds except few) conductor higher valent on-site substitutions fill the mobile holes and decreases both conductivity and superconductivity of the parent system. The examples are $Ba^{2+}$ site $La^{3+}$, $Sm^{3+}$, $Pr^{3+}$ and $Cu^{2+}$ site $Co^{3+}$, $Fe^{3+}$, $La^{3+}$, $Ru^{5+}$ substitutions in $YBa_2Cu_3O_7$ compound [1-3]. On the other hand in a p-type conductor the lower valent on-site substitutions, viz. $Y^{3+}$ site $Ca^{2+}$ increases the carriers and improves the superconducti vity of under-doped system [4,5]. This is however the most simplistic picture being given above. In reality with various alliovalent substitutions, the induction or reduction of carriers is accompanied with various structural changes and also the charge neutrality is not straight forward as the overall oxygen content of the system changes [6,7]. For example, in $Y_{1-x}Ca_xBa_2Cu_3O_7$ system, some of the carriers being introduced by $Y^{3+}/Ca^{2+}$ substitution are compensated by decrease in over all oxygen content of the system [6,7], which is not the case when parent system is under-doped viz. $Y_{1-x}Ca_xBa_2Cu_3O_{6.6}$ [4,5]. Hole filling by higher or hole creation by lower valent substitutions independently has been studied extensively over the years [4-7]. At the same time substitutional studies pertaining to simultaneous hole filling and hole creation in a composite HTSc system are still not explored fully [8,9]. Moreover in such a composite system viz. $Y_{1-x}Ca_xBa_2Cu_{3-y}Co_yO_7$, the structural changes with both hole creating ($Y^{3+}/Ca^{2+}$) and hole filling ($Cu^{2+}/Co^{3+}$) taking place simultaneously might be complicated. In this short article we report the phenomenon of hole filling and hole creation in a composite system $Y_{1-x}Ca_xBa_2Cu_{2.7}Co_{0.3}O_7$ compounds with x = 0.1 and 0.2. More interestingly we report the revival of Superconductivity by $Y^{3+}/Ca^{2+}$ substitution in $YBa_2Cu_{2.7}Co_{0.3}O_7$ without reported [9] phase transformation. The results of phase formation (XRD), resistivity ($\rho$), and thermo-electric power (S), are reported on $Y_{1-x}Ca_xBa_2Cu_{2.7}Co_{0.3}O_7$ compounds with x = 0.1 and 0.2.

**EXPERIMENTAL**

Samples of $Y_{1-x}Ca_xBa_2Cu_{2.7}Co_{0.3}O_7$ system with x = 0.1 and 0.2 synthesized by solid- state reaction route from ingredients of $Y_2O_3$, $CaCO_3$, $BaCO_3$, CuO and $Co_3O_4$. Calcinations were carried out on the mixed powder at 900, 910, 915 and 925$^0$C each for 24 hours with intermediate grindings. The pressed circular pellets were annealed in a flow of oxygen at 920$^0$C for 40 hours and subsequently cooled slowly to room temperature with an intervening annealing for 24 hours at 600$^0$C. X-ray diffraction (XRD) patterns were obtained at room temperature (MAC Science: MXP18VAHF$^{22}$; Cu$K_\alpha$ radiation). Resistivity Measurements were carried out by conventional four-probe method. Thermoelectric power (TEP) measurements were carried out by dc differential technique over a temperature range of 5 – 300 K, using home made set up. Temperature gradient of ~ 1 K is maintained throughout the measurement.

**RESULTS AND DISCUSSION**

Room temperature X-ray diffraction (XRD) patterns of $YBa_2Cu_3O_7$, $YBa_2Cu_{2.7}Co_{0.3}O_7$, and $Y_{0.8}Ca_{0.2}Ba_2Cu_{2.7}Co_{0.3}O_7$ are shown in Fig. 1. Pristine $YBa_2Cu_3O_7$ (Y:123) system is orthorhombic with *a* = 3.826(5) Å, *b* = 3.892(4) Å and *c* = 11.6734(7) Å.

In Y: 123 type compounds, a sheet of Cu and O atoms with variable composition $CuO_x$ interconnects the $BaO/CuO_2/Y/CuO_2/BaO$ slabs. The oxygen sites in Cu-$O_2$ planes are identified as O (2) and O (3). The O (2) resides between two Cu atoms along a-axis, while the one towards b-axis is named O (3) site. The copper atoms in Cu-$O_2$ planes are termed as Cu (2), while in $CuO_x$ chains are named as Cu (1). The oxygen site in Ba-O plane is named as O (4), while the RE-plane is found to be devoid of any oxygen. The oxygen sites in $CuO_x$ strings, often called as Cu-O chains, are named as O (5) (along a-axis) and O (1) (along b-axis) sites. In orthorhombic Y:123, O(1) are fully occupied, while O(5) are unoccupied, giving rise to b > a. With 10% Co doping at Cu-site i.e. $YBa_2Cu_{2.7}Co_{0.3}O_7$ both *a*, *b* lattice parameters become equal and system turns to be tetragonal with *a* = *b* = 3.856 (4)Å and *c* = 11.6637(8) Å. With Co substitution at Cu site in Y:123 the c-lattice parameter is decreased, due to lower ion $Co^{2+/3+}$ ion substitution at $Cu^{2+}$ site. This result is in agreement with previous reports on Cu/Co substitution [1-3,9]. The explanation for tetragonal structure is the occupation of otherwise un-filled O(5) sites in $CuO_x$ chains. The orthorhombic or the tetragonal structure of Y:123 system can be identified by looking at some characteristic peaks intensity and nature of splitting. For orthorhombic system, at 2(θ) of around $47.6^0$ and $48.9^0$ the split peaks appear with high intensity low angle and low intensity high angle sequence having indices [020], [200] and [123], [213] respectively, which is the case for $YBa_2Cu_3O_7$ sample, see, bottom XRD in Fig.1. As the orthorhombic distortion of the system decreases the split peaks start merging with each other. Interestingly when system becomes tetragonal the split peaks sequence of angle and intensity reverses. For tetragonal Y:123 though the sequence at 2(θ) of around $47.6^0$ and $48.9^0$ becomes low intensity low angle and high intensity high angle with indexing of [006], [200] and [116], [213], respectively, please see middle XRD in Fig.1 for $YBa_2Cu_{2.7}Co_{0.3}O_7$. When the splitting of these peaks is not clear, one has to deconvolute them and carry out the Reitveld analysis to confirm the structure. In present case as the splitting of these characteristic peaks is clear thus the need for deconvolution or the Reitveld analysis is not necessary. The XRD of $Y_{0.8}Ca_{0.2}Ba_2Cu_{2.7}Co_{0.3}O_7$ compound is shown in top part of Fig.1. As seen from the splitting nature characteristic peaks in the XRD, pattern the compound is tetragonal. According to some previous reports [9], $Y^{3+}/Ca^{2+}$ substitution in tetragonal $YBa_2Cu_{3-y}Co_yO_7$ system had brought about the tetragonal to orthorhombic phase transformation. On the contrary our XRD results clearly show that $Y_{0.8}Ca_{0.2}Ba_2Cu_{2.7}Co_{0.3}O_7$ compound is tetragonal.

With Co substitution at Cu site in Y:123 the *c*-lattice parameter is decreased, due to lower ion $Co^{2+/3+}$ ion substitution at $Cu^{2+}$ site [1]. The lattice parameters for x = 0.10 and 0.20 samples of series $Y_{1-x}Ca_xBa_2Cu_{2.7}Co_{0.3}O_7$ are respectively, *a* = *b* = 3.843(5)Å, c = 11.6691(9)Å, and *a* = *b* = 3.838(4)Å, c = 11.6703(8)Å respectively. With Ca substitution in $YBa_2Cu_{2.7}Co_{0.3}O_7$ *a* lattice parameter has a slight decreasing trend. This is due to the fact that though the *a* lattice parameter is supposed to increase slightly due to relatively bigger ion Ca substitution, the increasing number of carriers due to $Y^{3+}/Ca^{2+}$ substitution decrease the in plane Cu(2)–O(2) distance and hence the former effect is nullified. It is known that increasing p-type carriers in HTSC compounds increase the hybridization of the in-plane Cu(3d) and O(2p) orbitals resulting in a decrease both in Cu(2)-O(2) distance and the *a*-lattice parameter [10]. The *c*-lattice parameter of Co doped samples increase monotonically with increasing x, indicating successful substitution of $Y^{3+}$ by bigger ion $Ca^{2+}$. The ionic size of $Ca^{2+}$ in eight-fold coordination number is 1.12 Å, while that of $Y^{3+}$ in the same co-ordination is 1.02 Å. The system remains tetragonal over the whole range of doping (20% of $Ca^{2+}$ at $Y^{3+}$). Monotonic increase of *c*-lattice parameter with x in $Y_{1-x}Ca_xBa_2Cu_3O_{7-\delta}$ system guarantees the substitution of $Ca^{2+}$ at $Y^{3+}$ site in the same co-ordination number of eight [6,7].

Fig. 2 depicts the resistivity versus temperature (ρ vs. T) behavior of the $Y_{1-x}Ca_xBa_2Cu_{2.7}Co_{0.3}O_7$ compounds with x = 0.1 and 0.2. ρ (T) measurements showed superconducting transition temperature ($T_c$) to decrease from 90 K ($YBa_2Cu_3O_7$) to 33 K for $YBa_2Cu_{2.7}Co_{0.3}O_7$ which with further Ca substitution increases from 33 K to 53 K ($Y_{0.9}Ca_{0.1}Ba_2Cu_{2.7}Co_{0.3}O_7$) and 67 K for $Y_{0.8}Ca_{0.2}Ba_2Cu_{2.7}Co_{0.3}O_7$. $T_c$ decreases first with Cu-site Co substitution by hole-filling and later recovers by simultaneous hole creation by $Y^{3+}$ site $Ca^{2+}$ substitution. $\rho_{300K}$ is least for pristine $YBa_2Cu_3O_7$ sample and highest for $YBa_2Cu_{2.7}Co_{0.3}O_7$ sample. This shows that with $Co^{2+/3+}$ ion substitution at $Cu^{2+}$ the hole filling had taken place and hence the normal state resistivity is increased. With $Y^{3+}$ site $Ca^{2+}$ substitution in $YBa_2Cu_{2.7}Co_{0.3}O_7$ the $\rho_{300K}$ is decreased, suggesting the creation of mobile holes in the system. Normal state resistivity ($\rho_{300K}$) of the various samples is also listed in Table 1. Besides, the $T_c$ and $\rho_{300K}$ values, the normal state conduction is semiconductor-like for $YBa_2Cu_{2.7}Co_{0.3}O_7$ and it changes to metallic with further $Y^{3+}$ site $Ca^{2+}$ substitution. Interestingly though the $Y^{3+}$ site $Ca^{2+}$ substitution in under-doped (due to hole filling) $YBa_2Cu_{2.7}Co_{0.3}O_7$ compound revives superconductivity by increasing $T_c$, decreasing $\rho_{300K}$ and improving normal state conduction, but the same is not reached to the level of $YBa_2Cu_3O_7$. This shows that the hole filling ($Co^{2+/3+}$ ion substitution at $Cu^{2+}$) and hole creation ($Y^{3+}$ site $Ca^{2+}$ substitution) either do not compensate completely with each other, or there are other negative effects taking place simultaneously.

The results of thermoelectric power (S) measurements on $Y_{1-x}Ca_xBa_2Cu_{2.7}Co_{0.3}O_7$ with x = 0.0, 0.10, and 0.20 are shown in inset of Fig.(2). The value of S at room temperature (290 K) is found to be positive for all the samples, indicating them to be predominantly hole (p) type conductors. $S_{290 K}$ is least for $YBa_2Cu_3O_7$ (~ 6μV/K, plot not shown) and maximum (~ 30μV/K) for $YBa_2Cu_{2.7}Co_{0.3}O_7$. Further, the value of $S_{290 K}$ decreases with x for $Y_{1-x}Ca_xBa_2Cu_{2.7}Co_{0.3}O_7$. Implying that the number of mobile p-type carriers increase with increase in x for $Y_{1-x}Ca_xBa_2Cu_{2.7}Co_{0.3}O_7$ system. For strongly correlated systems the absolute value of S is reported to be inversely proportional to the number of mobile carriers [11]. However this may not be the fact in all the situations, and should be considered with reservations [12]. For example though in x = 0.20 sample of presently studied $Y_{1-x}Ca_xBa_2Cu_{2.7}Co_{0.3}O_7$ series, the overall number of carriers is more, but its $S_{290 K}$ value (~ 12μV/K) is also more than as for x = 0.10 (~ 10μV/K). Yet the general convention holds that decreasing value of S implicates for increase in mobile carriers. Further with decreasing temperature S passes through a maximum ($S_{max}$) and later starts decreasing with further decrease in temperature. The temperature corresponding to $S_{max}$ i.e. T ($S_{max}$) decreases monotonically with increasing x. Thermoelectric power measurements below T ($S_{max}$), exhibit transitions to $T_c^{S=0}$ at around 90 K, 29 K, 51 K, and 65 K respectively for $YBa_2Cu_3O_7$, $YBa_2Cu_{2.7}Co_{0.3}O_7$, $Y_{0.9}Ca_{0.1}Ba_2Cu_{2.7}Co_{0.3}O_7$ and $Y_{0.8}Ca_{0.2}Ba_2Cu_{2.7}Co_{0.3}O_7$ samples. In brief one can conclude that thermoelectric power measurements corroborate the resistance versus temperature results shown in Fig.2.

**CONCLUSION**

Results of phase formation, resistivity ($\rho$), and thermo-electric power (S), for $Y_{1-x}Ca_xBa_2Cu_{2.7}Co_{0.3}O_7$ compounds with x = 0.1 and 0.2, showed the Revival of Superconductivity by $Y^{3+}/Ca^{2+}$ substitution in $YBa_2Cu_{2.7}Co_{0.3}O_7$ without previously reported phase transformation.

**FIGURE CAPTIONS**

Fig.1 XRD patterns for $Y_{1-x}Ca_xBa_2Cu_{2.7}Co_{0.3}O_7$ system.

Fig.2 $\rho(T)$ for $Y_{1-x}Ca_xBa_2Cu_{2.7}Co_{0.3}O_7$ system, inset shows S(T) for the same.

Fig. 1 Ansari et al. (HU-10),

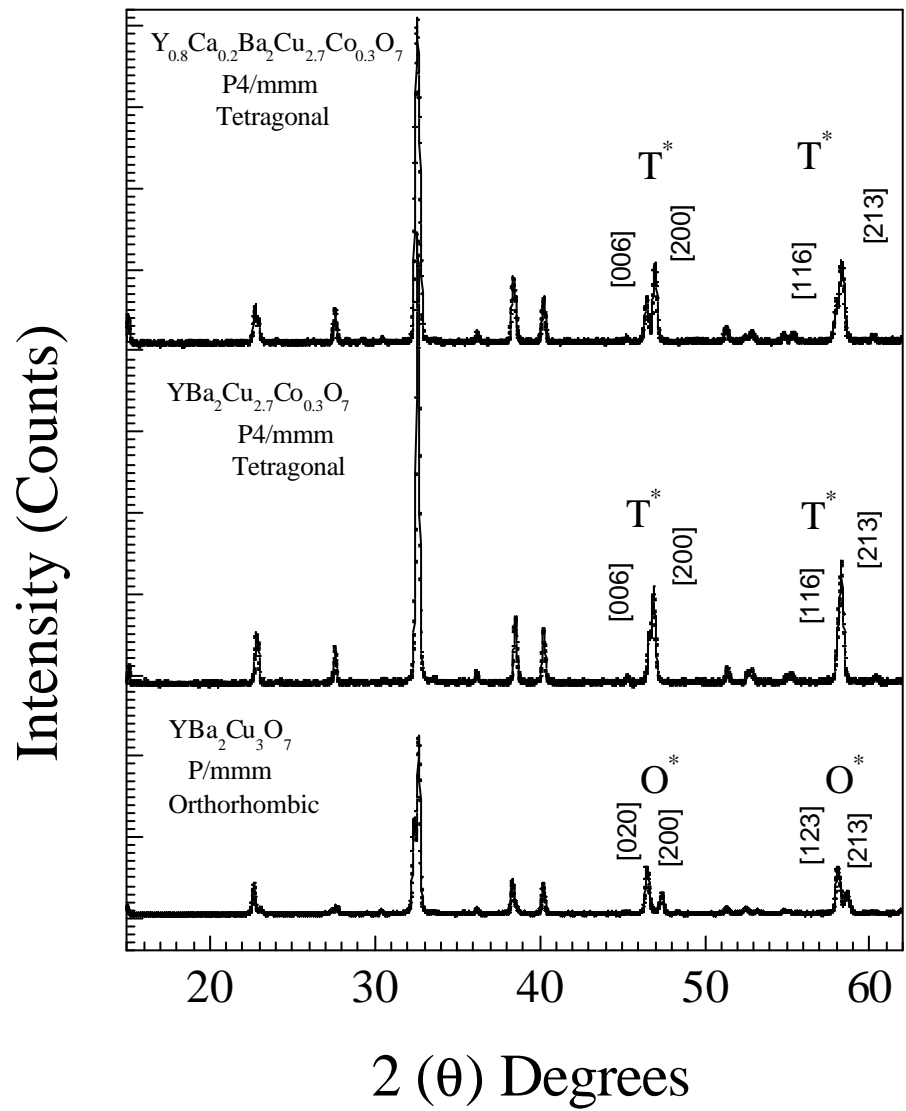

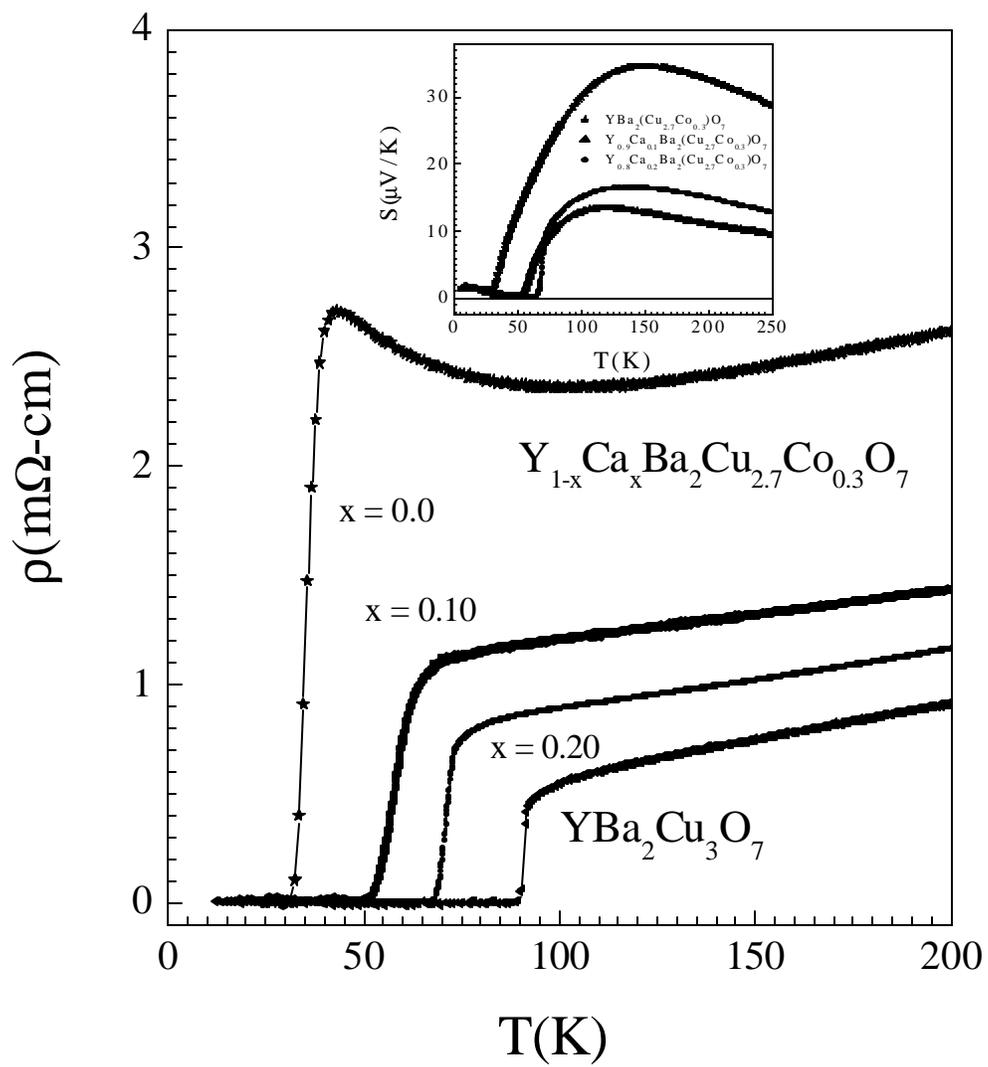